\documentclass{ws-procs975x65}

\begin{document}

\markboth{Gergely}
{Black holes on cosmological branes} 
%\wstoc{Black holes on cosmological branes}
%{Gergely}

\title{BLACK HOLES ON COSMOLOGICAL BRANES \footnote{
Research supported by OTKA grants no. T046939, TS044665 and the J\'{a}nos
Bolyai Fellowships of the Hungarian Academy of Sciences. The author wishes
to thank the organizers of the 11th Marcel Grossmann Meeting for support.} }
\author{L\'{A}SZL\'{O} \'{A}. GERGELY}

\address{Departments of Theoretical and Experimental Physics, 
University of Szeged,\\
D\'om t\'er 9, H-6720 Szeged, Hungary\\
\email{gergely@physx.u-szeged.hu}}

\begin{abstract}
While in general relativity black holes can be freely embedded into a
cosmological background, the same problem in brane-worlds is much more
cumbersome. We present here the results obtained so far in the explicit
constructions of such space-times. We also discuss gravitational collapse in
this context.
\end{abstract}


\bodymatter

\section*{}

Although almost perfectly homogeneous and isotropic at very large scales, as
probed by the measurements of the cosmic microwave background, our universe
contains local inhomogeneities in the form of galaxies and their clusters.
Therefore the cosmological model of Friedmann-Lema\^{\i}tre-Robertson-Walker%
\ (FLRW) geometry with flat spatial sections, considered valid on large
scales, has to be modified on lower scales. The simplest way to do it in
general relativity is to cut out spheres of constant comoving radius from
the FLRW space-time and fill them with Schwarzschild vacua, modeling stars,
black holes or even galaxies with a spherical distribution of dark matter.
Such a model was worked out by Einstein and Straus..\cite{ES} \ In the
framework of this, so-called Swiss-cheese model, it was shown that (a)
cosmic expansion has no influence on planetary orbits and (b) the
luminosity-redshift relation receives corrections \cite{Kantowski}. The
Einstein-Straus model however is unstable against perturbations.\cite%
{Krasinski}

Brane-world models\cite{ADD1,ADD2,RS1,RS2,MaartensLR} with our universe as a 
$4$-dimensional hypersurface (the brane) with tension $\lambda $ embedded in
a $5$-dimensional bulk is also confronted with the challenge of introducing
local inhomogeneities in the cosmological background. The basic dynamic
equation in these models is the effective Einstein equation,\cite{SMS,Decomp}
\begin{equation}
G_{ab}=-\Lambda g_{ab}+\kappa ^{2}T_{ab}+\widetilde{\kappa }^{4}S_{ab}-%
\overline{\mathcal{E}}_{ab}+\overline{L}_{ab}^{TF}+\overline{\mathcal{P}}%
_{ab}\ .  \label{modEgen}
\end{equation}%
On the right hand side we find the unconventional source terms $%
S_{ab}=[-T_{ac}^{\ }T_{b}^{c}+TT_{ab}/3-g_{ab}(-T_{cd}^{\
}T^{cd}+T^{2}/3)/2]/4$, quadratic in the energy-momentum tensor $T_{ab}$
(modifying early cosmology\cite{BDEL}); the average taken over the two sides
of the brane of the electric part $\mathcal{E}_{ab}=\widetilde{C}%
_{abcd}n^{b}n^{d}$ of the bulk Weyl tensor $\widetilde{C}_{abcd}$ (in a
cosmological context $\mathcal{E}_{ab}$ is known as dark radiation with
magnitude limited by Big Bang Nucleosynthesis (BBN) arguments\cite%
{BDEL,BBNconstraint}); the asymmetry source term $\overline{L}_{ab}^{TF}$
which is the trace-free part of the tensor $\overline{L}_{ab}=\overline{K}%
_{ab}\overline{K}-\overline{K}_{ac}\overline{K}_{b}^{c}-g_{ab}(\overline{K}%
^{2}-\overline{K}_{ab}\overline{K}^{ab})/2$ (with $\overline{K}_{ab}$ the
mean extrinsic curvature); and the pull-back to the brane $\mathcal{P}%
_{ab}=(2\widetilde{\kappa }^{2}/3)(g_{a}^{c}g_{b}^{d}\widetilde{\Pi }%
_{cd})^{TF}$of the bulk energy momentum tensor $\widetilde{\Pi }_{ab}$ (with 
$\kappa ^{2}$ and $\widetilde{\kappa }^{2}$ the brane and bulk coupling
constants and $g_{ab}$ the induced metric on the brane). The function $%
\Lambda =(\widetilde{\kappa }^{2}/2)(\lambda -n^{c}n^{d}\widetilde{\Pi }%
_{cd}-\overline{L}/4)$ contains the possibly varying normal projection of
the bulk energy-momentum tensor and the trace of the embedding function $%
\overline{L}_{ab}.$ Under special circumstances $\Lambda $ becomes the brane
cosmological constant. Here we consider this simpler case; also $\overline{%
\mathcal{E}}_{ab}=0=\overline{\mathcal{P}}_{ab}$.

For a perfect fluid with energy density $\rho $ and pressure $p$ the
non-linear source term $S_{ab}$ scales as $\rho /\lambda $ as compared to $%
T_{ab}$. Due to the huge value of the brane tension, this ratio is in
general infinitesimal, excepting the very early universe and the final
stages of gravitational collapse. The strongest bound on $\lambda $ was
derived by combining the results of table-top experiments on possible
deviations from Newton's law, probing gravity at sub-millimeter scales\cite%
{Gaccuracy} with the known value of the 4-dimensional Planck constant. In
the 2-brane model\cite{RS1} this gives\cite{Irradiated} $\lambda >138.59\,\,$%
TeV$^{4}.$Much milder limits arise from BBN constraints\cite{nucleosynthesis}
($\lambda \gtrsim 1$ MeV$^{4}$) and astrophysical considerations on brane
neutron stars\cite{GM} ($\lambda >5\,\times 10^{8}$ MeV$^{4}$).
Nevertheless, even when small, the presence of the source terms $S_{ab}$
implies that the pressure of the perfect fluid at the junction surface with
a vacuum region does not vanish.\cite{ND}.

The junction conditions between FLRW and Schwarzschild regions on the brane%
\cite{NoSwissCheese} imply a Swiss-cheese model that forever expands and
forever decelerates. The energy density and pressure of the fluid tend to
the general relativistic values at late times (on the physical branch; there
is also an unphysical branch never allowing for positive values of $\rho $).
At early times however $\rho $ is smaller than in the Einstein-Straus model
and $p$ takes large negative values.\cite{SwissCheese} When we allow for a
cosmological constant in the FLRW regions, the deviation from the
Einstein-Straus model is present at late-times as well. As the universe
expands, first $p$ turns positive, then eventually $\rho $ turns negative.
Moreover, for $\Lambda $ overpassing a threshold value $\Lambda _{\min \text{
}}$a pressure singularity accompanied by regular cosmological evolution
appears.

Such a Swiss-cheese model may be interpreted as a cosmological brane
penetrated by a collection of bulk black strings.\cite{SwissCheese} When the
brane is embedded asymmetrically, with different left and right bulk
regions, the source term $\overline{L}_{ab}^{TF}$ slightly modifies this
scenario.\cite{AsymSwissCheese} The evolution of the cosmological fluid is
further degenerated, proceeding along four possible branches, two of them
being physical. The future pressure singularity becomes generic, it appears
even below the threshold for $\Lambda $, due to the difference in the bulk
cosmological constants. For any $\Lambda <\Lambda _{\min }$\thinspace there
is a critical value of a suitably defined asymmetry parameter which
separates Swiss-cheese cosmologies with and without pressure singularities.%
\cite{AsymSwissCheese}

The mathematically similar problem of the gravitational collapse on the
brane has been also studied. If the pressure of the collapsing fluid is set
to zero, we recover the analogue of the general relativistic
Oppenheimer-Snyder collapse.\cite{OppSny} But in contrast with general
relativity, the exterior space-time is either characterized (beside the
mass) by a tidal charge\cite{tidalRN} (and the collapse possibly leads to a
bounce, a black hole or a naked singularity), or is non-static,\cite{GM,BGM}
infiltrated by radiation,\cite{DadhichGhosh,GovenderDadhich} or by a Hawking
flux.\cite{CasadioGermani}. An effective Schwarzschild solution on the brane
can be found when phantom bulk radiation is absorbed on the brane.\cite{Pal}

By allowing for non-vanishing pressure in the collapsing star, the exterior
can be again static.\cite{BraneOppSny1,BraneOppSny2} In this case the
collapsing fluid is described by the FLRW metric, which fills spheres of
constant comoving radius cut out from the Schwarzschild space-time. The
modified gravitational dynamics (\ref{modEgen}) again gives two branches. On
the physical branch the fluid is near dust-like at the beginning of the
collapse: it has an infinitesimal negative pressure (tension) $p=w\rho $
with $w\approx -\rho /2\lambda $, arising from the interaction of the fluid
with the brane. The tension vanishes in the general relativistic limit, but
as the collapse proceeds and $\rho $ increases, it becomes more important.
For astrophysical brane black holes the tension stays small even at horizon
forming. However well below the horizon, at the final stages of the collapse 
$w\approx -1/2$ and the condition for dark energy $\rho +3p<0$ is obeyed.
This however has little repulsive effect, as at such high energy densities
the source term $S_{ab}$ (which is always positive) dominates, and the
singularity inevitably forms.


\begin{thebibliography}{99}
\bibitem{ES} A.\! Einstein and\! E.\! G.\! Straus,\! Rev.\! Mod.\! Phys.\! 
\textbf{17},\! 120\! (1945), errata,\! ibid.\textit{\ }\textbf{18},\! 148\!
(1946).

\bibitem{Kantowski} R. Kantowski, Astrophys.\! J. \textbf{155},\! 89\!
(1969).

\bibitem{Krasinski} A.\! Krasi\'{n}ski,\! Inhomogeneous Cosmological\!
Models,\! Cambridge\! University\! Press\! (1997).

\bibitem{ADD1} N. Arkani-Hamed, S. Dimopoulos, and G. Dvali, Phys. Lett. B 
\textbf{429}, 263 (1998).

\bibitem{ADD2} N. Arkani-Hamed, S. Dimopoulos, and G. Dvali, Phys. Rev. D 
\textbf{59}, 086004 (1999).

\bibitem{RS1} L. Randall and R. Sundrum, Phys. Rev. Lett. \textbf{83}, 3370
(1999).

\bibitem{RS2} L. Randall and R. Sundrum, Phys. Rev. Lett. \textbf{83}, 4690
(1999).

\bibitem{MaartensLR} R. Maartens R, Living Rev. Rel. \textbf{7}, 1 (2004).

\bibitem{SMS} T. Shiromizu T, K. Maeda, and M. Sasaki, Phys. Rev. D \textbf{%
62}, 024012 (2000).

\bibitem{Decomp} L. \'{A}. Gergely, Phys. Rev. D \textbf{68}, 124011 (2003).

\bibitem{BDEL} P. Bin\'{e}truy, C. Deffayet, U. Ellwanger, and D. Langlois,
Phys.Lett.\textit{\ }B \textbf{477}, 285 (2000).

\bibitem{BBNconstraint} K.Ichiki,\! M.Yahiro,\! T.Kajino,\! M.Orito, \!\!
and\! G.J.Mathews,\! Phys.Rev.D\textbf{66},\! 043521\! (2002).

\bibitem{Gaccuracy} J. C. Long, et al., Nature \textbf{421}, 922 (2003).

\bibitem{Irradiated} L. \'{A}. Gergely and Z. Keresztes, JCAP\textbf{\ 06}%
(01), 022 (2006).

\bibitem{nucleosynthesis} R.Maartens,\! D.Wands,\! B.A.Bassett,\! and\!
I.P.C.Heard, \! Phys.Rev.D\textbf{62},\! 041301(R)\! (2000).

\bibitem{GM} C. Germani and R. Maartens, Phys. Rev. D \textbf{64}, 124010
(2001).

\bibitem{ND} N. Deruelle, gr-qc/0111065 (2001).

\bibitem{NoSwissCheese} L. \'{A}. Gergely, Phys. Rev. D \textbf{71}, 084017
(2005), erratum, ibid. \textbf{72}, 069902 (2005).

\bibitem{SwissCheese} L. \'{A}. Gergely, Phys. Rev. D \textbf{74}, 024002
(2006).

\bibitem{AsymSwissCheese} L. \'{A}. Gergely, I. K\'{e}p\'{\i}r\'{o},
hep-th/0608195 (2006).

\bibitem{OppSny} J. R. Oppenheimer and H. Snyder, Phys. Rev. \textbf{56},
455 (1939).

\bibitem{tidalRN} N.Dadhich,\! R.Maartens,\! P.Papadopoulos, \! and\!
V.Rezania, Phys.Lett.B\textbf{487},\! 1\! (2000).

\bibitem{BGM} M. Bruni, C. Germani, and R. Maartens, Phys. Rev. Lett.\textit{%
\ }\textbf{87}, 231302 (2001).

\bibitem{DadhichGhosh} N. Dadhich N and S. G. Ghosh, Phys. Lett.\textit{\ }B 
\textbf{518}, 1 (2001).

\bibitem{GovenderDadhich} N. Dadhich and S. G. Ghost, Phys. Lett. B \textbf{%
538}, 233 (2002).

\bibitem{CasadioGermani} R. Casadio and G. Germani, Prog. Theor. Phys. 
\textbf{114}, 23 (2005).

\bibitem{Pal} S Pal, Phys. Rev. D 74 124019 (2006).

\bibitem{BraneOppSny1} L. \'{A}. Gergely, hep-th/0603254, JCAP \textbf{07}%
(02), 027 (2007).

\bibitem{BraneOppSny2} L. \'{A}. Gergely, gr-qc/0606073, to appear in Int.
J. Mod. Phys. D (2006).
\end{thebibliography}
\end{document}